# Predictive Gain Estimation – A mathematical analysis

P.Chakrabarti
Sir Padampat Singhania University
Udaipur , Rajasthan , India
Email_id : prasun9999@rediffmail.com

*Abstract-* **In case of realization of successful business, gain analysis is essential. In this paper we have cited some new techniques of gain expectation on the basis of neural property of perceptron. Support rule and Sequence mining based artificial intelligence oriented practices have also been done in this context. In the view of above fuzzy and statistical based gain sensing is also pointed out.**

## I. INTRODUCTION

Predictive gain estimation plays a pivotal role in forecast based strategic business planning. Realization of gain patterns is essential for proper gain utilization. the gain analysis can be sensed using perceptron learning rule and other techniques. Prediction can also been performed on the basis of fuzzy assumption theory on the factors on which gain of a business organization depends. Statistical application of gain estimation can also be used as a tool in this context.

## II. GAIN ANALYSIS USING PERCEPTRON LEARNING RULE

We assume that $G = \{g_1, g_2, \ldots, g_n\}$ be the set of gain of business organization noted after observation in respective years $Y = \{y_1, y_2, \ldots y_n\}$. As per our assumption $g_k$ ( where $g_k \in G$ and $1 < k < n$).

The following steps must be done:

1. Achieve difference in gain estimate as $\delta_i = | g_{i+1} - g_i |$, (i=1 to n-1).

2. $g_k$ is optimum i.e. $\text{Max}(g_1, g_2, \ldots, g_n) = g_k$

3. Compute
$$\delta_{AVG} = \sum_{i=1}^{n-1} \delta_i / (n-1)$$

4. Predicted gain = $g_n = g_{n-1} + \delta_{AVG}$ ( if $g_{n-1} < \delta_{AVG}$ )
or $g_n = g_{n-1} - \delta_{AVG}$ ( if $g_{n-1} > \delta_{AVG}$ )

5. Normalization of the parameters on which gain depends will be scaled upwards/downwards by a factor of x where $x = (g_n - g_k)$.

## III. BEHAVIORAL THEORY OF GAIN ESTIMATION

For a particular business, yearwise gain is noted. It is possible to rationalize this sequence of gain patterns on the basis of likelihood estimate , support rule and sequence mining.

### A. Likelihood measure

In case of gain prediction of a business, forecast based expectation can be achieved after sensing original gains.

**Theorem1 :** Maximum likelihood estimator of expected gain , taking on threshold , depends on individual estimated gain.

Proof : Let $g_1, g_2, \ldots, g_n$ be the gain of business organization noted after observation in respective years $y_1, y_2, \ldots y_n$. The business authority initially expects gain as $g_e$.
The joint density function of the gain patterns for that particular business is given by
$L = f(g_1, g_e) f(g_2, g_e) \ldots f(g_n, g_e)$ where L is the likelihood. We can write as follows:
$\delta f(g_1, g_e)/\{f(g_1, g_e).\delta g_e\} + \ldots \delta f(g_n, g_e)/\{f(g_n, g_e).\delta g_e\} = 0 \ldots..(1)$
Solution of $g_e$ from eq(1) reveals that expected gain depends on individual estimated gain.

### B. Gain analysis based on support rule

The factors on which gain of a particular business depends are mainly production , quality , market competition , risk involvement and cost and the above parameters are denoted as P,Q,M,R,C respectively. Let ,
PL = low production ,PH = high production,QM= medium quality,QB= best quality,ML= low market competition, MH= high market competition, RL= less risk involvement, RH= high risk involvement,CL= low cost, CH= high cost

| YEAR | GAIN | PARAMETER STATUS |
|------|------|------------------|
| $y_1$ | $g_1$ | PL  QB  MH  RL  CH |
| $y_2$ | $g_2$ | PH  QM  ML  RL  CL |
| $y_3$ | $g_3$ | PH  QB  MH  RH  CH |
| $y_4$ | $g_4$ | PL  QM  MH  RH  CH |

Table 1 : Gain observation with parameter status

As per support rule, support count is as follows:
PL = PH = 1/2
QM=QB= 1/2
MH =3/4
ML= 1/4
RL=RH=1/2
CH=3/4
CL=1/4

Optimum Gain can be achieved if the gain condition = $g_e$ = f(PH,QB,MH,RL,CL).

*C. Gain analysis based on sequence mining*

Sequence of gain patterns of a business organizarion depends on Boolean notation of its associated factors.
PL=0 ,PH=1, QM=0, QB=1,
ML=0,MH=1,RL=0,RH=1,CL=0 and CH=1.
As per Table 1, the sequence tables are as follows:

| GAIN | P | Q | M | R | C |
|------|---|---|---|---|---|
| $g_1$ | 0 | 1 | 1 | 0 | 1 |
| $g_2$ | 1 | 0 | 0 | 0 | 0 |
| $g_3$ | 1 | 1 | 1 | 1 | 1 |
| $g_4$ | 0 | 0 | 1 | 1 | 1 |

Table 2 : Sequence observation with parameter status

As per the sequence, it is quite obvious that the market competition is high and hence there must be ample production with high cost and best quality.

## IV. CHAOS THEORY BASED GAIN ESTIMATION

The chaos theory can be applied in case of gain estimation. As per the theory, a minute change in the input will affect the output a lot. Hence if a business organization bestows gain in an uniform rate, and due to some external factor, huge gain/loss is incurred, then inspection of the internal parameter status can be observed using Chaos Theory. Factors involved are production, quality, market competition, risk involvement and cost. By crisp operation, status of each has two probable outputs – 0 or 1. The bit values expected are X={0,1} and the combinations are $2^5$.

## V. FUZZY BASED OPTIMUM GAIN REALIZATION

If the gain estimates for a particular business are as $g_1$, $g_2$, $g_3$ and $g_4$, then the codes in terms of fuzzy value are as follows:- $g_1$ = 0.1, $g_2$ = 0.2, $g_3$ = 0.3, $g_4$ = 0.4. We assume that (i)the outcomes of the gains are based on parameters P,Q,M,R,C as before (ii)the parameter P will be high in cases of the gains $g_2$, $g_3$ ; Q will be on $g_1$, $g_3$ ; M on $g_1$, $g_3$, $g_4$, R on $g_1$, $g_2$ and C on $g_2$.

The desired optimum gain will be a fuzzy set whose member functions will be such that is intersection rule is applied in case of R and C while for other parameters the union rule.

| Statement | Fuzzy Values |
|-----------|--------------|
| $g_1$ | 0.1 |
| $g_2$ | 0.2 |
| $g_3$ | 0.3 |
| $g_4$ | 0.4 |

Table 3: Fuzzy based value assignment

The above chart involves the gains and its respective fuzzy values and set. Since the gains are based on fuzzy operation, so we can denote each gain as a fuzzy set and the results as per our assumption are shown below: -

| Fuzzy Set | Member Function | Value |
|-----------|-----------------|-------|
| $\tilde{G_1}$ | P,Q,M,R,C | $\tilde{G_1}$= {($x_1$, 0.1), ($x_2$, 0.6), ($x_3$, 0.9), ($x_4$, 0.2), ($x_5$, 0.8)} |
| $\tilde{G_2}$ | | $\tilde{G_2}$= {($x_1$, 0.7), ($x_2$, 0.3), ($x_3$, 0.5), ($x_4$, 0.3), ($x_5$, 0.2)} |
| $\tilde{G_3}$ | | $\tilde{G_3}$= {($x_1$, 0.8), ($x_2$, 0.8), ($x_3$, 0.7), ($x_4$, 0.7), ($x_5$, 0.7)} |
| $\tilde{G_4}$ | | $\tilde{G_4}$= {($x_1$, 0.2), ($x_2$, 0.1), ($x_3$, 0.8), ($x_4$, 0.9), ($x_5$, 0.8)} |

Table 4: Fuzzy representation of respective gains

Now, $\mu_{(\tilde{G_1} \cup \tilde{G_2} \cup \tilde{G_3} \cup \tilde{G_4})}(x_1)$ = max(0.1, 0.7,0.8,0.2) = 0.8;
$\mu_{(\tilde{G_1} \cup \tilde{G_2} \cup \tilde{G_3} \cup \tilde{G_4})}(x_2)$ = max(0.6, 0.3,0.8,0.1) =0.8;
$\mu_{(\tilde{G_1} \cup \tilde{G_2} \cup \tilde{G_3} \cup \tilde{G_4})}(x_3)$ = max(0.9, 0.5,0.7,0.8) =0.8;
$\mu_{(\tilde{G_1} \cap \tilde{G_2} \cap \tilde{G_3} \cap \tilde{G_4})}(x_4)$ = min(0.2, 0.3,0.7,0.9) = 0.2;
$\mu_{(\tilde{G_1} \cap \tilde{G_2} \cap \tilde{G_3} \cap \tilde{G_4})}(x_5)$ = min(0.8, 0.2,0.7,0.8) = 0.2

Hence $\tilde{G_{opt}}$ = {($x_1$, 0.8), ($x_2$, 0.8), ($x_3$, 0.8), ($x_4$,0.2),($x_5$,0.2)}

| PARAMETER | YEAR OF REALIZATION |
|-----------|---------------------|
| P | $y_3$ |
| Q | $y_3$ |
| M | $y_1$ |
| R | $y_1$ |
| C | $y_2$ |

Table 5: Corresponding year realization for optimum gain

In this case, we realize that the business has flourished in year $y_1$ maintaining lowest risk in highest competitive market ; lowest cost involvement in $y_2$ while highest production with best quality in $y_3$.

## VI. STATISTICAL APPROACH OF GAIN PREDICTION

We can analyze gain in time series form. In a business, the year of highest gain is inspected for future use and prediction.

**Theorem 2**
If a gain of a particular business changes (G) over time (t) in an exponential manner, in that case the value of the gain at the centre point an interval $(a_1, a_2)$ is a geometric mean of its values at $a_1$ and $a_2$.

Proof: Let $G_a = mn^a$, m and n being constants.
Then $G_{a1} = mn^{a1}$ and $G_{a2} = mn^{a2}$
Now, value of G at $(a_1 + a_2)/2$
$$= mn^{(a1+a2)/2}$$
$$= [m^2 n^{(a1+a2)}]^{1/2}$$
$$= [(mn^{a1})(mn^{a2})]^{1/2}$$
$$= (G_{a1} G_{a2})^{1/2}$$

**Theorem 3**
If a variable m representing gain of a business is related to another variable n representing year in the form m = an, where a is a constant, then harmonic mean of n is related to that of n based on the same equation.

Proof: Let x is number of observed gain values.
If $m_{HM} = x / (\sum 1/m_i)$ for i = 1 to x
$= x / (\sum 1/an_i)$ [ Since $m_i = an_i$ ]
$= x / (1/a \sum 1/n_i)$ for i = 1 to x
$= a( x / (\sum 1/n_i))$ for i = 1 to x
$= an_{HM}$

Gain can also be estimated based on probabilistic approach. Suppose G be the set of gains $g_1, g_2, g_3 \ldots g_m$ for a particular business organization with respective probability $p_1, p_2, \ldots p_n$.

When $\sum_{i=1}^{m} p_i = 1$

then $E(G) = \sum_{i=1}^{m} g_i p_i = 1 \ldots\ldots\ldots\ldots\ldots\ldots(2)$

provided it is finite.

Here, we are use bivariate probability based on G $(g_1, g_2, g_3 \ldots g_m)$ i.e. set of achieved gains and Q $(q_1, q_2, q_3, \ldots q_n)$ i.e. set of predictive gains, ( 1 < m < n)

**Theorem 4**
If the values of observed gain and predicted gain be two jointly distributed random variables then
$E(G + Q) = E(G) + E(Q)$.

Proof : G assume values $g_1, g_2, g_3 \ldots\ldots g_m$
Q assume values $q_1, q_2, q_3 \ldots\ldots q_m$

$P(G=g_i, Q = q_j) = p_{ij}$, i = 1 to n and j = 1 to n

$E(G + Q) = \sum_i \sum_j (g_i + q_j) p_{ij}$

$= \sum_i \sum_j g_i p_{ij} + \sum_i \sum_j q_j p_{ij}$

$= \sum_i g_i \sum_j p_{ij} + \sum_j q_j \sum_i p_{ij}$

$E(G + Q) = E(G) + E(Q) \ldots\ldots\ldots\ldots\ldots\ldots(3)$

We can also predict gain on the basis of autoregression property
Here from next year onwards, a future gain guessing will be done based on

$g_{m+1} = \varphi + \phi_m g_m + \phi_{m-1} g_{m-1} + \phi_{m-2} g_{m-2} + \ldots + \ldots \phi_1 k_g + \omega_{m+1} \ldots\ldots(4)$

Here, $\omega_{m+1}$, indicates a random error at time m+1. Here, each element in the time series can be viewed as a combination of a random error and a linear combination of previous values. Here $\phi_i$ are the autoregressive parameters.

We can also predict based on the theory of moving average. In this case the gain guessing strategy will be based on :

$g_{m+1} = a_{m+1} + \theta_m a_m + \theta_{m-1} a_{m-1} + \theta_{m-2} a_{m-2} + \ldots\ldots + \theta_{m-q} a_{m-q} \ldots(5)$
where $a_i$ is a shock
$\theta$ is estimate
q is term indicating last predicted value.

The values of the gain estimates for different years are not identical. In some cases the values are close to one another, where in some cases they are highly dedicated from one another. In order to get a proper idea about overall nature of a given set of values, it is necessary to know, besides average, the extent to which the gain estimates differ among themselves or equivalently, how they are scattered about the average. Let the values $g_1, g_2, g_3 \ldots\ldots g_m$ are the gain estimates values and c be the average of the original values of $g_{m+1}, g_{m+2}, \ldots\ldots g_n$.

Mean Deviation of k about c will be given by

$MD_c = \dfrac{1}{(n-m)} \sum_{i=1}^{n-m} | g_i - c | \ldots\ldots\ldots\ldots\ldots\ldots(6)$

In particular, when $c = \overline{g}$, mean deviation about mean will be given by

$MD_{\overline{g}} = \dfrac{1}{(n-m)} \sum_{i=1}^{n-m} | g_i - \overline{g_i} | \ldots\ldots\ldots\ldots\ldots(7)$

## VII. CONCLUSION

The paper deals with several gain estimates which play crucial role in strategic business planning thereby bestowing optimum gain estimate. In this context gain analysis with neural , fuzzy and statistical justification pointed out in this paper.

About author:

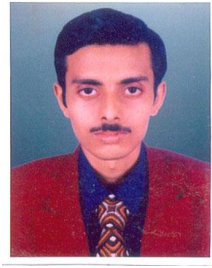

Dr.P.Chakrabarti(09/03/81) is currently serving as Associate Professor in the department of Computer Science and Engineering of Sir Padampat Singhania University,Udaipur. Previously he worked at Bengal Institute of Technology and Management , Oriental Institute of Science and Technology, Dr.B.C.Roy Engineering College, Heritage Institute of Technology, Sammilani College. He obtained his Ph.D(Engg) degree from Jadavpur University in Sep09,did M.E. in Computer Science and Engineering in 2005,Executive MBA in 2008and B.Tech in Computer Science and Engineering in 2003.He is a member of Indian Science Congress Association , Calcutta Mathematical Society , Calcutta Statistical Association , Indian Society for Technical Education , Computer Society of India, VLSI Society of India , Cryptology Research Society of India, IEEE(USA) , IAENG(Hong Kong) ,CSTA(USA) and Senior Member of IACSIT(Singapore) .He is a Reviewer of International journal of Information Processing and Management(Elsevier), International Journal of Engineering and Technology, Singapore. He has about 90 papers in national and international journals and conferences in his credit. He has several visiting assignments at BHU Varanasi , IIT Kharagpur , Amity University,Kolkata , et al.